# Frequency-angular distribution for terahertz emission of single-color laser filament plasma under electrostatic field

L. Seleznev,[1, a)] G. Rizaev,[1] D. Pushkarev,[1] A. Koribut,[1] Yu. Gerasimova,[1] Ya. Grudtsyn,[1] S. Savinov,[1] Yu. Mityagin,[1] D. Mokrousova,[1] and A. Ionin[1]

*P. N. Lebedev Physical Institute of RAS, 53 Leninski pr., Moscow, 119991 Russia*

(Dated: 25th March 2021)

Frequency-angular distributions for THz emission in the range of 0.1 – 3 THz generated in a single-color laser filament plasma both under and without electrostatic field are for the first time experimentally studied. The angular distribution for various spectral components of this THz emission is demonstrated to differ significantly. The maximal propagation angles for these components grow up under the electric field. The angular distribution for the low-frequency THz emission depends significantly on the laser pulse energy in contrast to the case of no electric field. An increase in the laser pulse energy leads to a decrease of the propagation angles for low-frequency THz emission and to disappearance of the local minimum in the angular distribution on the propagation axis.

---

a)Electronic mail: seleznev@lebedev.ru

Terahertz emission within the range of 0.1 – 10 THz[1,2] is quite promising for various applications in biology[3,4], diagnostics of drugs and microchips[5], in security systems[6], medicine[7], in the study of art objects[8,9] and other fields. Many books and reviews are devoted to generation of THz emission and its applications (see, for example, Refs. 10 and 11). One of the THz emission sources is air plasma formed during filamentation of ultrashort laser pulses[12]. This method of producing THz emission is attractive and may be used to build a laboratory broadband THz source based on a commercial femtosecond laser. THz emission formed by a single-color laser filament plasma was demonstrated in Refs. 13 and 14 to propagate forward into a cone with a minimum on the optical axis (a hollow cone). There was demonstrated in Refs. 15 and 16 that, if the filament plasma was placed under an external electrostatic field, the THz signal amplitude increased considerably and the angular distribution changed to a cone entirely filled by THz emission with a maximum on the axis. However, to the best of our knowledge, the angular distribution of THz spectrum for the laser filament plasma has never been presented including Refs. 15 and 16. The object of our paper is experimental study of THz emission angular distribution within the millimeter and submillimeter spectral range of 0.1 – 3 THz for different spectral components generated in plasma channel of a single-color laser filament both under the external electrostatic field and in its absence.

In the experiments we use 100 fs pulses of Ti:Sa laser system with the central wavelength of 740 nm. The laser beam diameter is 3 mm (FW1/eM). Laser pulse energy reaches 3 mJ and can be varied by diffractive attenuators. Laser pulses are focused by a lens with the focal distance of 30 cm, so that the geometric focus coincides with the back edge of the flat electrodes (Fig. 1a) to ensure for the plasma channel position to be just within the interelectrode gap. The electrodes length along the propagation axis and the interelectrode gap is 15 mm and 5 mm, respectively, the applied voltage being up to several kilovolts. Figure 1a shows the position of the electrodes for the horizontal direction of the electric field. (In the experiments, we also observed an angular distribution of THz emission for the vertical direction of the electric field — horizontal position of the electrodes. The change in the field direction did not lead to significant changes in the angular distribution of THz emission). The lens focal point is located on the rotation axis around which a superconducting hot-electron NbN bolometer is rotated. The bolometer can detect THz radiation in the range from 0.1 to 6 THz. To measure angular distributions of different THz spectral components, narrow-band THz filters are placed directly in front of the bolometer window. Spectral characteristics of these filters are shown in Fig. 1b.

Fig. 2a demonstrates the angular distribution of different THz spectral components generated in a single-color laser filament plasma without external electric field. The laser pulse energy in this experiment is 2 mJ. The angular distribution of the THz low-frequency component of 0.1 THz corresponds to a hollow cone distribution with a local minimum on the propagation axis, THz emission being observed under wide angles up to 25 degrees relative to the propagation axis. High-frequency components of the THz emission (0.3 – 1.0 THz) do not have a pronounced minimum and propagate at significantly smaller — less than 15 degrees — angles, the higher frequency components being within a narrower cone. It should be noted that at frequencies of 3 THz and higher no signal is detected. For rough estimates of the total angular distribution of the wide-frequency THz pulse, we summarize the observed signals for all the spectral components. In Fig. 2a it corresponds to hollow circles. The total THz signal turns out to be the hollow cone, as it was observed in the previous paper 13.

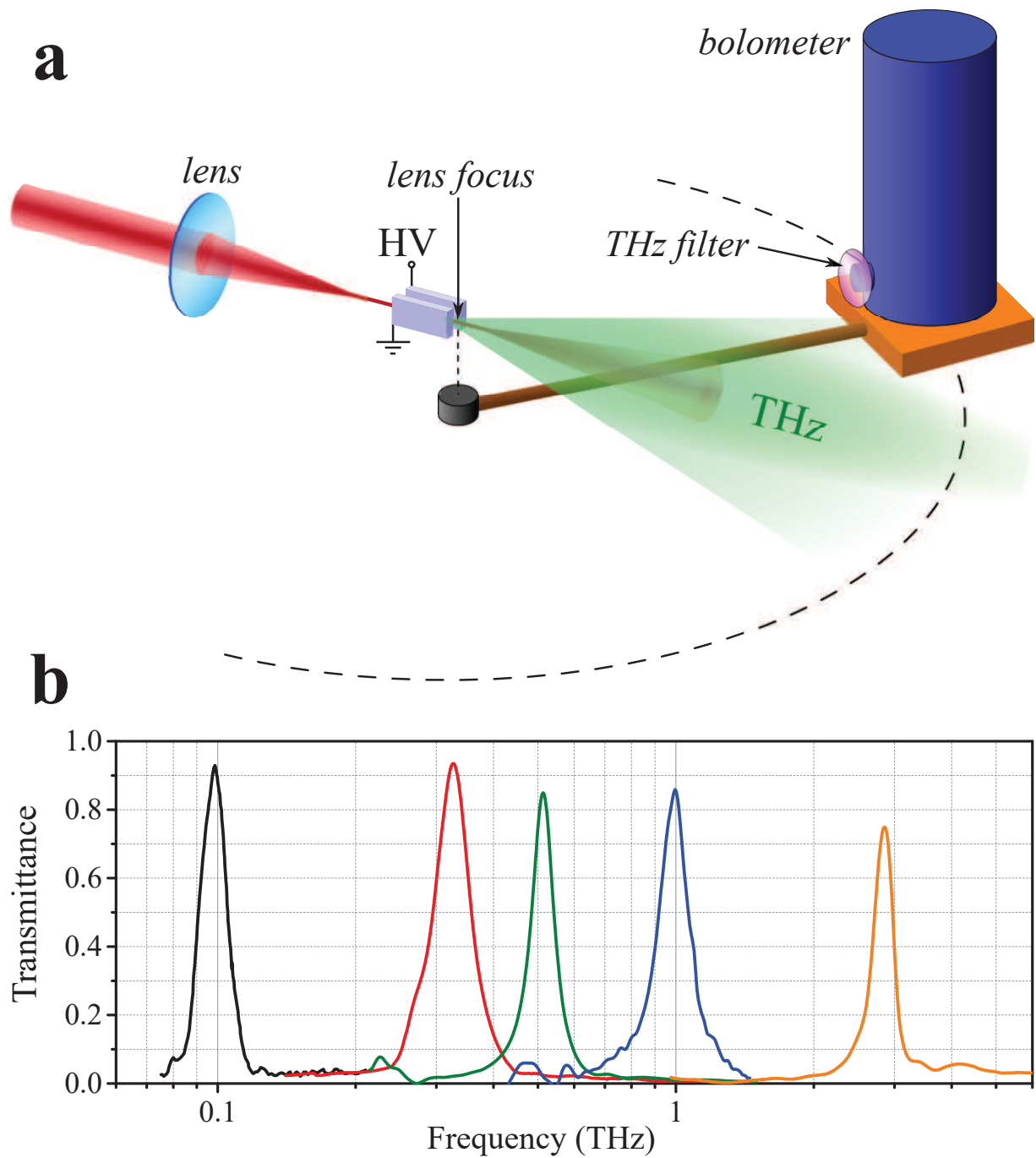

Figure 1. Experimental scheme (a) and spectral characteristics of narrowband THz filters (b).

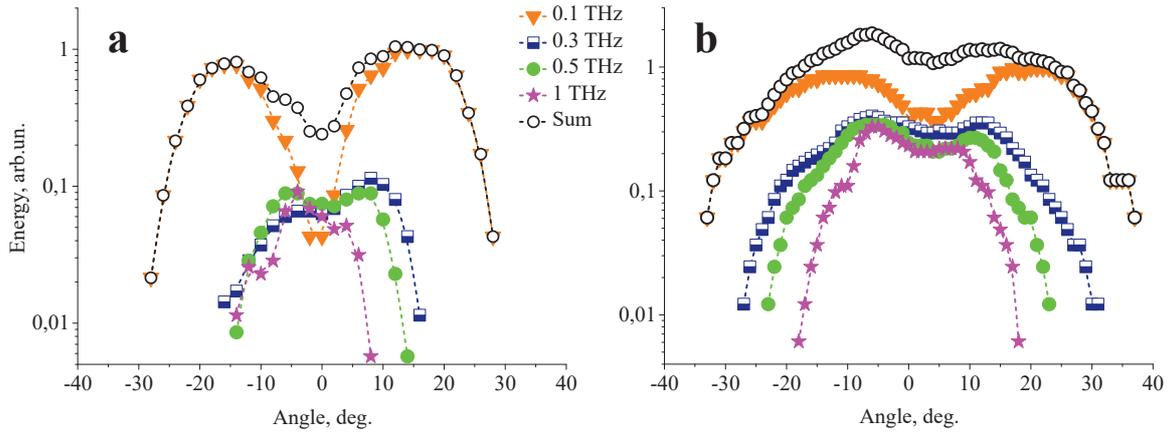

Figure 2. Angular distribution of different spectral THz components without (a) and with (b) external field of 5 kV/cm. The laser pulse energy is 2 mJ.

The angular distributions of different THz spectral components under external field of 5 kV/cm are presented in Fig. 2b. Under these conditions, low-frequency emission of 0.1 THz also propagates in a hollow cone, the dip on the axis diminishing and propagation angles expanding. In contrast to the low-frequency components, the high-frequency ones propagate quite closer to the axis. One can notice a clear trend of the cone narrowing with THz frequency growth. The maximal propagation angle of low-frequency 0.1 THz, 0.3 THz and 1 THz components reach 40, 30 and 20 degrees, respectively. At the same time, summarizing signals of all measured spectral components, we obtain an entirely filled cone of THz emission (empty circles in Fig.2b), that is also in good agreement with the previously observed results in Refs. 15 and 16.

An increase in the laser pulse energy leads to an elongation of the laser filament plasma channel and a slight growth of the linear plasma density, which practically does not affect the total angular distribution of the wide-frequency THz radiation[14]. Fig. 3a,b demonstrates the angular distribution obtained with the filters for 0.1 (a) and 1 THz (b) for various laser pulse energies without external

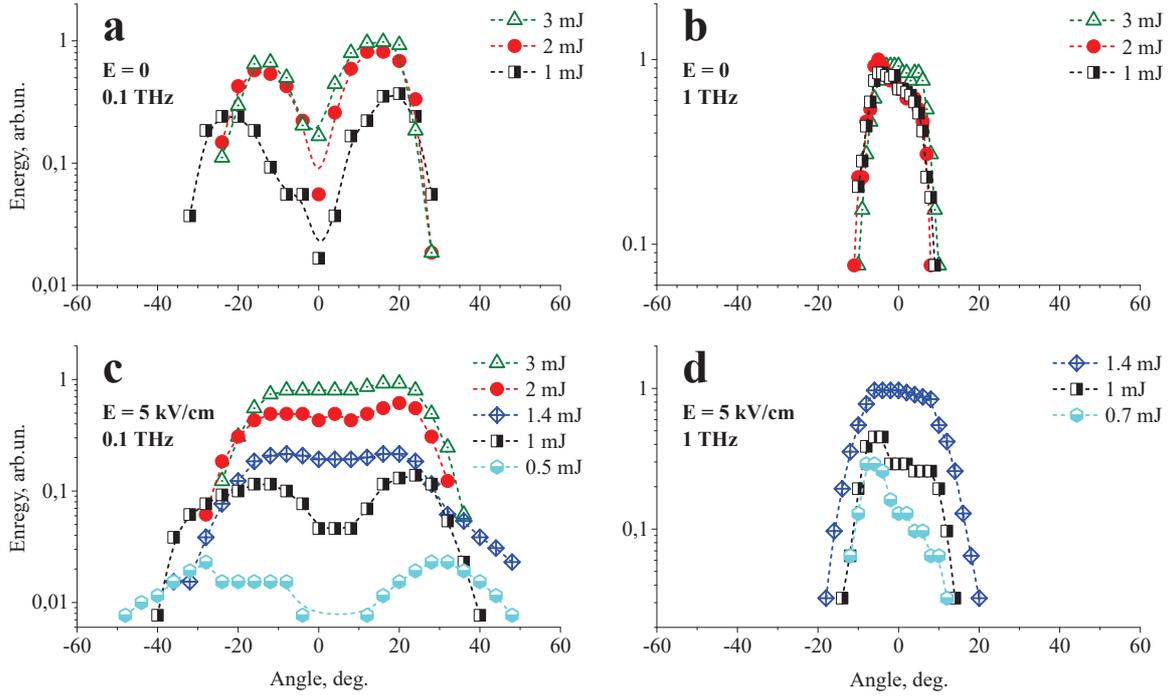

Figure 3. Angular distributions of THz emission: (a) 0.1 THz without external field, (b) 1 THz without field, (c) 0.1 THz at field strength 5 kV/cm, (d) 1 THz at field strength 5 kV/cm

field. It is worth noting that for 1 THz frequency the growth of the laser pulse energy not only retains the angular distribution, but also practically does not change the signal amplitude. In contrast to 1 THz emission, an increase in the laser pulse energy in the low-frequency region of 0.1 THz leads to the THz cone slight narrowing and the signal amplitudes growth. However, the angular distribution type with a local minimum on the propagation axis does not change.

In the presence of the external electric field, the energy of the laser pulse significantly influences the THz emission angular distribution in the low-frequency region (Fig. 3c). In the experiments, while the energy increases from 0.5 mJ to 3 mJ, the maximal propagation angles go down from 50 to 30 degrees. In case of relatively low energy, we observe a clear hollow cone with a pronounced dip on the axis. An increase in laser pulse energy leads to the dip decrease down to the complete absence of a local minimum. Also, the detected THz signal rises by almost two orders of magnitude at the same strength of the electric field. In the high-frequency spectral region (Fig. 3d), with a growth of the laser pulse energy, the type of the distribution shape remains practically unchanged, but, in contrast to the case of no field, the amplitude of THz signals increases.

In conclusion, we for the first time experimentally studied the frequency-angular distribution of THz radiation generated in a single-color laser filament plasma both without and under the external electrostatic field. It is shown that without the electric field the angular distribution of various THz spectral components differs significantly. While for the low-frequency components of 0.1 THz, a 'classical' conical distribution of the THz emission with a minimum on the axis is observed, the high-frequency components of 0.3 – 1 THz propagate into a narrower cone, which has practically no minimum on the axis. The high-frequency THz components propagate within the narrower cone, the angle of which does not depend on the laser pulse energy. When the external electric field is applied, the maximal propagation angles for THz spectral components grow up. Unlike the case of no electric field, the angular distribution of the low-frequency THz emission depends significantly on the laser pulse energy. The laser pulse energy growth leads to a decrease of the propagation angles for low-frequency THz components and to disappearance of the local minimum in the angular distribution on the propagation axis.

The research is supported by RFBR grant 20-02-00114.